\documentclass{ws-procs975x65}
\usepackage{latexsym}
\usepackage{caption}
\usepackage{amssymb}

\begin{document}
\title{Toward an invariant definition of repulsive gravity}

\author{Orlando Luongo and Hernando Quevedo}
\address{Dipartimento di Fisica, Universit\`a di Roma "La Sapienza", Piazzale Aldo Moro 5, I-00185 Roma, Italy.\\
ICRANet and ICRA, Piazza della Repubblica 10, I-65122 Pescara,
Italy.}

\begin{abstract}
A remarkable property of naked singularities in general relativity
is their repulsive nature. The effects generated by repulsive
gravity are usually investigated by analyzing the trajectories of
test particles which move in the effective potential of a naked
singularity. This method is, however, coordinate and observer
dependent. We propose to use the properties of the Riemann tensor in
order to establish in an invariant manner the regions where
repulsive gravity plays a dominant role. In particular, we show that
in the case of the Kerr-Newman singularity and its special subcases
the method delivers plausible results.
\end{abstract}

\section{Introduction}
It is well known that the field equations of Einstein's theory of gravity
allow the existence of exact solutions containing naked singularities. Moreover,
recent studies  indicate that under certain circumstances
naked singularities can appear as the result of
a realistic gravitational collapse \cite{joshi07}.
An intriguing property of many naked singularities is that
they can generate repulsive gravity.
To understand this repulsive nature one can study the motion
of test particles which, for example in the case of
stationary axially symmetric
fields, reduces to the study of an effective potential. Although the explicit
form of the effective potential depends  on the type of motion under consideration, in general one can find
certain similitudes between the effective potential for geodesic motion and the effective Newtonian potential which
follows from the metric as $g_{tt} \approx 1 - 2 V_{N} = 1 - 2 M_{eff}/r$, where the effective mass reduces to the physical mass
$M$ at infinity. One can then
intuitively expect that in the regions where $M_{eff}$ becomes negative, the effects of repulsive gravity may occur.
In the case of the Schwarzschild metric the effective mass
coincides with the physical mass, and repulsive gravity is obtained only if we change $M\rightarrow-M$; hence, the source
of repulsion can be considered as unphysical. However, in the cases of the Reissner-Nordstr\"om and Kerr metrics
we have respectively $M_{eff}=M-\frac{Q}{2r}$ and $M_{eff}=M-L(a,r,\theta)$,\cite{pap66} leading to  spacetime regions
where repulsive gravity exists. The disadvantage of this approach is that it is clearly coordinate and observer dependent.
The attempts to define gravitational repulsion in terms of curvature invariants \cite{def89} and the behavior
of light cones \cite{def08} are also not definite.
In this work we propose to use the eigenvalues of the curvature tensor to characterize repulsive gravity
in an invariant manner. We first consider the main second order curvature invariants and show that they do not
reproduce the simple case of the Schwarzschild naked singularity. Then we show that the curvature eigenvalues
provide a reasonable solution to the problem.

\section{An invariant approach}
>From the curvature tensor one can form 14  functionally
independent scalars of which only 4 are non-zero in empty space
\cite{deb56}. As for the second order invariants, the most
interesting are the Kretschmann scalar,
$K_1=R_{\alpha\beta\gamma\delta}R^{\alpha\beta\gamma\delta}$, the
Chern-Pontryagin scalar, $K_2 =
[{}^*R{}]_{\alpha\beta\gamma\delta}R^{\alpha\beta\gamma\delta},$ and
the Euler scalar $K_3 =
[{}^*R{}^*]{}_{\alpha\beta\gamma\delta}R^{\alpha\beta\gamma\delta}$,
where the asterisk represents dual conjugation. Although the use of
these invariants has been proposed to define ``repulsive domains"
and negative effective masses in curved spacetimes \cite{def89},
their quadratic structure does not allow to consider all possible
cases of naked singularities. Indeed, for the Schwarzschild
spacetime we get $K_1=48M^2/r^6$, whereas $K_2$ and $K_3$ are
proportional to $K_1$. Since the change $M\rightarrow -M$ does not
affect the behavior of $K_1$, these invariants do not recognize the
presence of a Schwarzschild naked singularity. Similar difficulties
appear in more general cases like the Kerr and Kerr-Newman naked
singularities. Therefore, it seems necessary to consider the only
first order invariant which is the curvature scalar $R$; however, it
vanishes identically in the empty space of naked singularities.

As an alternative approach we propose to use the eigenvalues of the curvature. To this end, consider the $SO(3,C)-$representation
of the curvature as follows. Let the line element be written in an (pseudo-)orthonormal frame as $ds^2 = \eta_{ab}\vartheta^a\otimes\vartheta^b$
with $\eta_{ab}={\rm diag}(+1,-1,-1,-1)$. From the curvature 2-form $\Omega^a_{\ b} = d\omega^a_{\ b} + \omega^a_{\ c}\wedge\omega^c_{_b}=\frac{1}{2}
R^a_{\ bcd} \vartheta^c\wedge\vartheta^d$, where $d\vartheta^a = -\omega^a_{\ b}\wedge \vartheta^b$, one obtains the components of the curvature
tensor whose irreducible parts are: the Weyl tensor,
$W_{abcd}=R_{abcd}+2\eta_{[a|[c}R_{d]|b]}+\frac{1}{6} R\eta_{a[d}\eta_{c]b}$, the trace-free Ricci tensor,
$ E_{abcd} = 2\eta_{[b|[c}R_{d]|a]} - \frac{1}{2}R\eta_{a[d}\eta_{c]b}$, and the curvature scalar, $S_{abcd} = -\frac{1}{6} R\eta_{a[d}\eta_{c]b}$,
with $R_{ab}=\eta^{cd}R_{cabd}$. Furthermore, using the bivector notation for the indices $ab\rightarrow A$, according to
$01\rightarrow 1,\ 02\rightarrow 2,\ 03\rightarrow 3,\ 23\rightarrow 4,\ 31\rightarrow 5,\ 12\rightarrow 6,$ the curvature tensor can be written as
$R_{AB} = W_{AB} + E_{AB} + S_{AB}$ with
\begin{equation}\label{dkjhf}
W_{AB}=\left(
         \begin{array}{cc}
           N & M \\
           M & -N \\
         \end{array}
       \right), E_{AB}=\left(
         \begin{array}{cc}
           P & Q \\
           Q & -P \\
         \end{array}
       \right),S_{AB}=\frac{R}{12}\left(
         \begin{array}{cc}
           I_{3} & 0 \\
           0 & I_{3} \\
         \end{array}
       \right).
\end{equation}
Here $M$, $N$ and $P$ are $(3\times 3)$ real symmetric matrices, whereas $Q$ is antisymmetric. The $SO(3,C)-$representation corresponds
to ${\bf R}=W+E+S$ with $W=M+iN$, $E=P+iQ$, and $S=\frac{1}{12} R\,I_3$ (see \cite{que92} for more details.) The eigenvalues of the curvature
matrix
${\bf R}$ are in general complex $\lambda_n=a_n+ib_n$ and, according to Petrov's
classification \cite{ES}, are an invariant characterization of the curvature tensor. Moreover, in the most general case of gravitational
fields belonging to Petrov's class I, we obtain the largest number of eigenvalues, namely $n=3$.

In the special case of the Schwarzschild metric there is
only one  eigenvalue $\lambda= M/r^3$ and the change $M\rightarrow -M$ induces a drastic change in the eigenvalue and in the
structure of spacetime as well. An analysis of the more general Kerr-Newman naked singularity indicates that in fact the curvature eigenvalues
change their sign and present several maxima and minima in the vicinity of the singularity which is exactly the region where repulsive gravity
appears. It then seems reasonable to introduce the concept of region of repulsion as the region of spacetime contained between the
first extremum of the eigenvalue, when approaching from spatial infinity,  and the singularity. The extremum is defined in an invariant manner
as $\partial \lambda_n/\partial x^i =0$, where $x^i$ are the spatial coordinates. This invariant approach leads to the following values for
the Reissner-Nordstr\"om and Kerr naked singularities
\begin{equation}
R^{^{RN}}_{rep} = 2\frac{Q^2}{M}\ ,\qquad R_{rep}^{^K} = \left(1+\sqrt{2}\right)a\cos\theta \ ,
\end{equation}
respectively. These results are in agreement with the analysis of
test particles. In fact, the Reissner-Nordstr\"om singularity
presents repulsion effects outside the classical radius $R_{class}
=Q^2/M$, and the radius of repulsion $R^{^{RN}}_{rep} = 2 R_{class}$
is always situated within the zone of instability of circular
motion. The Kerr naked singularity turns out to be attractive only
on the equatorial plane $[R_{rep}^{^K}(\pi/2) =0$], and it is
repulsive otherwise. The case of the Kerr-Newman singularity cannot
be solved analytically in a compact form. On the axis, however, the
radius of repulsion is given by the largest root of the equation $
Mr^4-2Q^2r^3-6Ma^2r^2+2a^2Q^2r+Ma^4=0$. Introducing values for the
mass, charge and angular momentum the resulting radius of repulsion
is always situated in the region where the motion of test particles
is affected by repulsive gravity.

Our invariant approach to define repulsive gravity leads to plausible and physical reasonable results in the case of naked singularities which
possess a black hole counterpart. The investigation of naked singularities generated by a mass quadrupole moment (without black hole counterparts)
indicates that our method consistently delivers the expected results. Moreover, it turns out that the concept of region of repulsion can be
used as a criterion  to study the problem of matching interior and exterior solutions of Einstein's equations.

\section*{Acknowledgements}

The authors want to thank prof. Roy Patrick Kerr and prof. Remo
Ruffini for discussions.

\end{document}